# Ultrawide Frequency Tuning of Atomic Layer van der Waals Heterostructure Electromechanical Resonators

Fan Ye[1†], Arnob Islam[1†], Teng Zhang[3,4], Philip X.-L. Feng[1,2*]

[1]*Department of Electrical, Computer, & Systems Engineering, Case School of Engineering, Case Western Reserve University, Cleveland, OH 44106, USA*

[2]*Department of Electrical & Computer Engineering, Herbert Wertheim College of Engineering, University of Florida, Gainesville, FL 32611, USA*

[3]*Department of Mechanical & Aerospace Engineering, and* [4]*BioInspired Syracuse, Syracuse University, Syracuse, NY 13244, USA*

## *Abstract*

We report on the experimental demonstration of atomically thin molybdenum disulfide ($MoS_2$)-graphene van der Waals (vdW) heterostructure nanoelectromechanical resonators with ultrawide frequency tuning. With direct electrostatic gate tuning, these vdW resonators exhibit exceptional tunability, in general, $\Delta f/f_0 >$200%, for *continuously tuning the same device and the same mode* (*e.g.*, from ~23 to ~107MHz), up to $\Delta f/f_0 \approx$370%, the largest fractional tuning range in such resonators to date. This remarkable electromechanical resonance tuning is investigated by two different analytical models and finite element simulations. Further, we carefully perform clear control experiments and simulations to elucidate the difference in frequency tuning between heterostructure and single-material resonators. At a given initial strain level, the tuning range depends on the two-dimensional (2D) Young's moduli of the constitutive crystals; devices built on materials with lower 2D moduli show wider tuning ranges. This study exemplifies that vdW heterostructure resonators can retain unconventionally broad, continuous tuning, which is promising for voltage-controlled, tunable nanosystems.

***Keywords***: van der Waals heterostructure, 2D materials, resonator, frequency tuning

[†]Equally Contributed Authors. [*]Corresponding Author. Email: philip.feng@ufl.edu.

Continuous frequency tuning and control over broad ranges via simply applying voltage is an indispensable function in state-of-the-art nano/microelectromechanical systems (N/MEMS) exploiting resonance modes, especially in oscillators, filters, mixers, and their applications [1,2,3]. To realize wide frequency tuning in resonant N/MEMS, device structures are engineered to enable high electromechanical coupling, and thus their resonance frequencies can be efficiently configured and controlled by electrical voltage or field [4,5,6,7,8]. In addition to desirable material properties and appropriate tuning methods, device structures also play a significant role in achieving excellent performance and diverse functionality. For example, by depositing a piezoelectric layer on top of a SiC beam, the resulting heterostructure beam (bimorph) could be tuned with the piezoelectric effect [9]. Although these strategies and traditional tunable N/MEMS have witnessed significant success in communication and sensing applications, the trends and challenges confronting the conventional scaling and miniaturization of electronic devices impose a great demand on exploring novel NEMS with unconventional materials, structures, and characteristics, such as highly efficient frequency tuning with ultrawide ranges.

The rise of atomically thin two-dimensional (2D) materials [10,11] and their van der Waals (vdW) heterostructures [12], endowed with ultralow bending stiffnesses and ultrahigh strain limits, offer exciting opportunities for building highly tunable heterostructure NEMS resonators and oscillators. For example, suspended $MoS_2$-graphene heterostructures exhibit robust resonances in the high and very high frequency (HF & VHF) bands [13]. Whereas initial efforts have been made on studying basic resonance characteristics [13,14,15] and moderate tunability [14,15] in such early devices, a clear understanding of frequency tuning in vdW heterostructures is still lacking. Specifically, there are several obvious open questions on this topic. First, the frequency tuning ability of vdW heterostructures is not fully explored. Second, there is no analytical or finite element method (FEM) model available to explain the tuning behavior of vdW heterostructure resonators. Third, a clear comparison of tuning capabilities between NEMS made of single 2D materials and those of vdW heterostructures is needed but is still lacking. Importantly, according to recent reports [14,15], interlayer slip between $MoS_2$ and graphene layers could occur presumably, even under a moderate applied gate voltage ($V_G$) (~2.3V to 8.3V), which might limit the applications of vdW heterostructure resonators. All these unknowns may inevitably result in an underestimation of the potential of vdW heterostructure NEMS. Therefore, a dedicated, comprehensive investigation and clear analysis of frequency tuning in vdW heterostructure resonators are highly desirable and needed.

In this work, using an all-dry, clean fabrication approach [16], we build atomically thin $MoS_2$-graphene vdW heterostructure resonators and investigate their frequency tuning by changing the applied gate voltage ($V_G$). We find that all the devices exhibit *continuous*, ultrabroad frequency tuning with $\Delta f/f_0 > 200\%$, with the largest tuning range up to $\Delta f/f_0 \approx 370\%$, which is much wider than reported data hitherto in electrostatically tuned resonators made of single 2D materials [17,18,19,20,21]. Unlike in previous reports [14,15], we observe no "kink" in the frequency tuning curves for any devices, even over a much wider applied voltage range, suggesting clean interfaces and strong vdW bonding. In parallel to the frequency tuning measurement, we employ two different analytical models and FEM simulations to reveal the governing mechanisms in tuning vdW heterostructure resonators. We further perform a clear control experiment to compare the tuning behaviors of vdW heterostructure and single 2D material resonators. With solid experiments and comprehensive, clarifying results, this study suggests a promising pathway toward building highly tunable vdW heterostructures that are important for voltage-controlled 2D resonators and oscillators.

## Results and Discussions

**Figure 1**a illustrates the scheme of a MoS$_2$-graphene heterostructure NEMS resonator with electrostatic excitation and tuning. The MoS$_2$-graphene heterostructures are fabricated using the completely dry-transfer method [16]. We stack single-layer (1L) MoS$_2$ on top of single-layer graphene (1LGr), tri-layer graphene (3LGr), and few-layer graphene (FLGr), respectively, to build 1LMoS$_2$-1LGr, 1LMoS$_2$-3LGr, and 1LMoS$_2$-FLGr vdW heterostructure drumhead resonators with a diameter ~3.2µm and a microcavity depth of 290nm. After device fabrication, a DC voltage ($V_G$) applied to the gate deflects the heterostructure membrane and efficiently varies the tension in the membrane. A small radio frequency (RF) voltage ($v_{RF}$) signal is superposed to the $V_G$ to excite the time-varying resonance motion. The resonance motion is detected by using a laser optical interferometry system integrated with Raman spectroscopy and photoluminescence (PL) measurement (Figure 1b) [5,13].

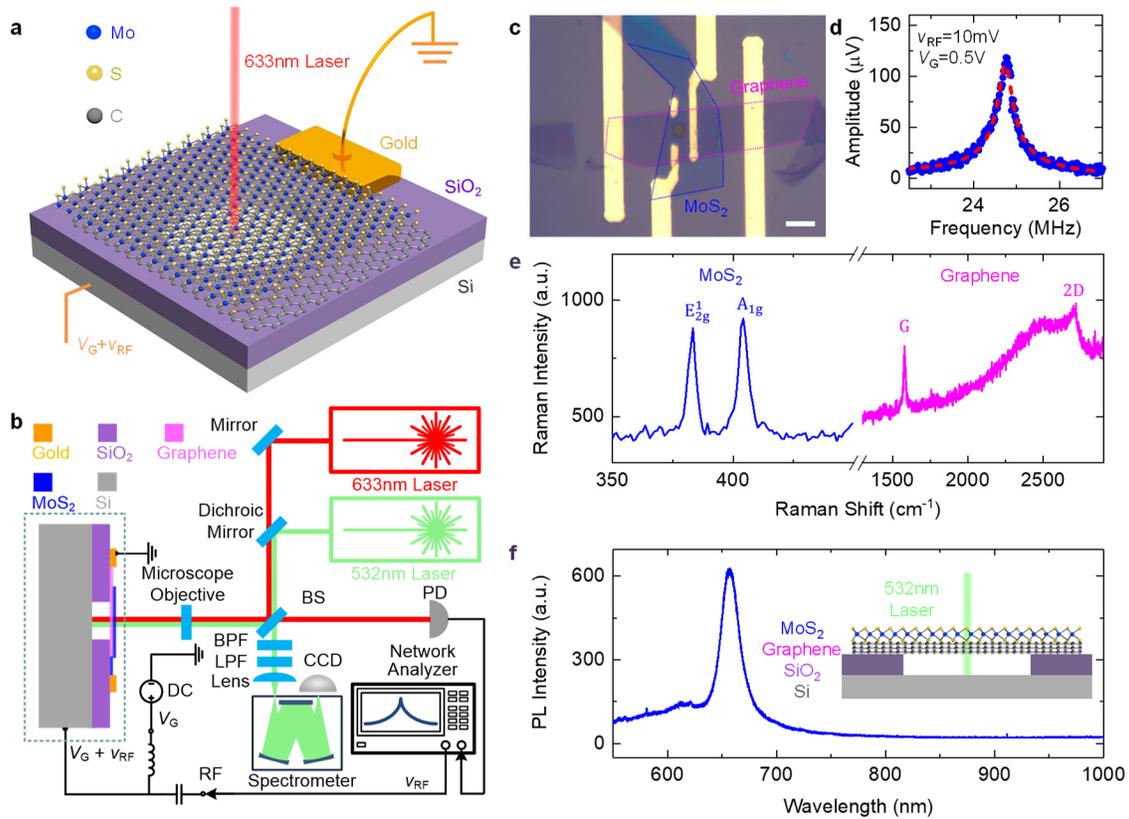

**Figure 1**: MoS$_2$-graphene van der Waals drumhead heterostructure resonators. (a) Illustration of a circular drumhead MoS$_2$-graphene van der Waals heterostructure resonator. The blue, yellow, and gray spheres represent molybdenum (Mo), sulfur (S), and carbon (C) atoms, respectively. (b) The laser optical interferometry measurement system. PD and BS represent photodetector and beam splitter, respectively. All measurements are performed in a moderate vacuum (~20mTorr) at room temperature. (c) Optical microscopy image of the 1LMoS$_2$-3LGr vdW heterostructure resonator. *Scale bar*: 10µm. (d) Fundamental-mode resonance of the 1LMoS$_2$-3LGr van der Waals heterostructure in (c). (e) Raman and (f) PL signatures measured from the suspended region of the device in (c). *Inset*: illustration of the Raman and PL measurement.

The optical microscopy image of the 1LMoS$_2$-3LGr heterostructure resonator is shown in Figure 1c. The 3L graphene flake (red dashed lines) is first transferred onto the substrate followed by another 1L MoS$_2$ flake (blue dashed lines) stacking on top of the graphene. This 1LMoS$_2$-3LGr device shows a clear resonance in moderate vacuum, with a fundamental frequency $f_0 \approx 24.7$MHz and quality factor $Q \approx 77$, indicating that the vdW membrane is continuous and suspended after stacking process (Figure 1d). Raman measurements are performed using a 532nm laser to characterize the number of layers of the graphene and MoS$_2$. From the Raman measurement results (Figure 1e), four characteristic peaks from the heterostructure (two MoS$_2$ peaks and two graphene peaks) are observed, confirming the suspended region is continuous MoS$_2$-graphene heterostructure. The number of layers of MoS$_2$ is verified by the separation between the $E_{2g}^1$ and $A_{1g}$ peaks [22], and the graphene thickness is verified by the intensity ratio between the G and 2D modes [23]. Photoluminescence (PL) measurements are performed (Figure 1f), and on each device we observe a clear direct bandgap peak without other indirect bandgap peaks, confirming the MoS$_2$ flake is single layer [11].

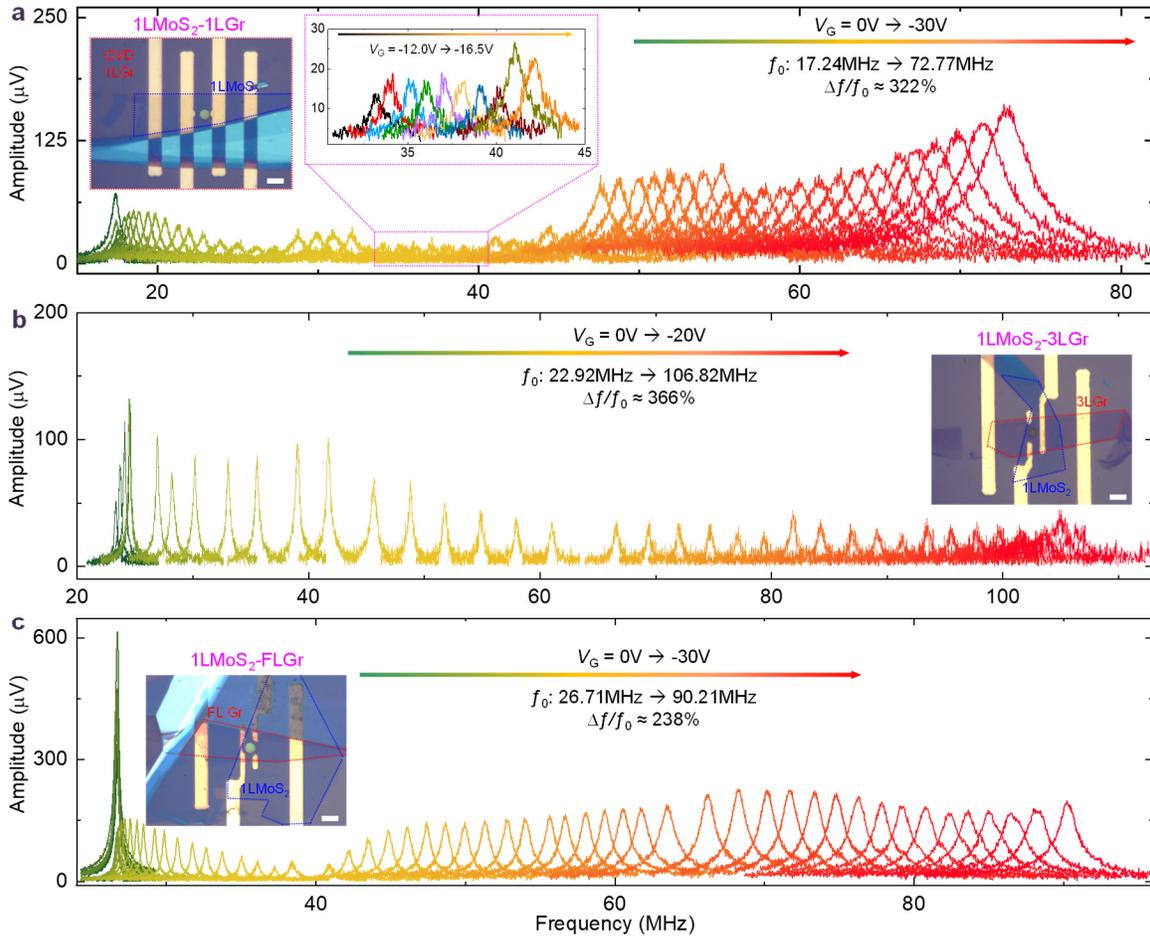

**Figure 2**: Ultrawide electrostatic frequency tuning measured from the 1LMoS$_2$-1LGr, 1LMoS$_2$-3LGr, and 1LMoS$_2$-FLGr nanoelectromechanical resonators as |$V_G$| increases. *Inset images*: optical microscopy images of the corresponding devices in (a), (b), and (c), respectively. *Inset plot* in (a): zoom-in view of the resonance mode for the 1LMoS$_2$-1LGr device with |$V_G$| increasing from 12.0V to 16.5V. *Scale bars*: 5μm.

We then investigate electrostatic frequency tuning by gradually increasing |$V_G$|. **Figure 2**a-c show an overview of frequency tuning of the 1LMoS$_2$-1LGr, 1LMoS$_2$-3LGr, and 1LMoS$_2$-FLGr

devices, respectively. As $V_G$ varies from 0 to -30V, the frequency upshifts from 17.24MHz to 72.77MHz ($\Delta f/f_0 \approx 322.1\%$) for the 1LMoS$_2$-1LGr heterostructure device. Similarly, the resonance frequencies increase from 22.92MHz to 106.82MHz ($\Delta f/f_0 \approx 366.1\%$) for the 1LMoS$_2$-3LGr device, and from 26.71MHz to 90.21MHz ($\Delta f/f_0 \approx 237.7\%$) for the 1LMoS$_2$-FLGr device, as $|V_G|$ is increased from 0 to 20V, and from 0 to 30V, respectively. This level of remarkably broad frequency tuning via simple electrostatic gating is unprecedented; and the tuning ranges are much wider than the highest $\Delta f/f_0$ values achieved by electrostatically tuned resonators based on conventional NEMS/MEMS materials (~4%) and even wider than the highest $\Delta f/f_0$ values attained by electrostatically tuned carbon nanotube resonators (200%) [4,24,25,26].

To analyze the observed ultrawide frequency tuning in the vdW heterostructure resonators, the $f_{res}$ versus $V_G$ data are plotted in **Figure 3**. Three devices exhibit continuous frequency tuning behavior with increasing applied gate voltage $V_G$, *i.e.*, a moderate frequency tuning in low voltage regime (around $|V_G| < 10$V for 1LMoS$_2$-1LGr and 1LMoS$_2$-3LGr devices, and $|V_G| < 15$V for 1LMoS$_2$-FLGr device) and increasingly wider tuning ranges as $|V_G|$ increases. It should be noted that we have not observed any measurable "kinks" described in previous reports [14,15] and attributed to interfacial slippage or other nonideal effects therein, suggesting clean and tightly bonded MoS$_2$-graphene interfaces in our vdW devices. To attain quantitative analysis and understanding, we employ two analytical models (see the *Supporting Information* for detailed equation derivations of these analytical models) and an FEM model by employing COMSOL Multiphysics simulations. The resonance frequency of the circular vdW heterostructure drumhead device under electrostatic gating is given by

$$f_{res} = (1/2\pi)\sqrt{k_{eff}/M_{eff}}, \qquad (1)$$

where $k_{eff}$ is the effective stiffness (spring constant) and $M_{eff}$ is the effective mass of the heterostructure membrane. In analytical model #1 (green dashed lines in Figure 3a,c,e), the stiffness $k_{eff}$ is obtained by the second-order differentiation of the total potential energy (sum of the electrostatic energy, $U_{electrostatic}$, and the stored elastic energy, $U_{elastic}$) $\dfrac{\partial^2(U_{electrostatic}+U_{elastic})}{\partial z^2}$, where $z$ is membrane equilibrium position under the DC gate polarization voltage, $V_G$. The frequency tuning based on analytical model #1 is given as [27]

$$f_{res} = \frac{1}{2\pi}\sqrt{\frac{4.9 E_{Y,Hetero} t_{Hetero} \varepsilon_0 - \dfrac{\epsilon_0 \pi a^2}{3d^3}V_G^2 + \dfrac{\pi \epsilon_0^2}{8(1-\nu^2)E_{Y,Hetero} t_{Hetero} \varepsilon_0^2}\dfrac{a^2}{d^4}V_G^4}{M_{eff}}}, \qquad (2)$$

where $E_{Y,Hetero}$, $\nu$, $\varepsilon_0$, $t_{Hetero}$, $d$, and $a$ are Young's modulus (conventional 3D definition, with unit [N/m$^2$] or [Pa]), Poisson's ratio, initial strain, device thickness, cavity depth, and radius of the device, respectively, $\epsilon_0$ is the vacuum permittivity, and $M_{eff} = \eta\pi a^2 t_{Hetero}\rho$ is the effective modal mass, with $\eta$ being a numerical constant set by the mode shape. Here we note the initial tension (with unit [N/m]) is $\gamma_0 = E_{Y,Hetero}\cdot t_{Hetero}\cdot\varepsilon_0 = E_{Y,Hetero,2D}\cdot\varepsilon_0$, where $E_{Y,Hetero,2D} = E_{Y,Hetero}\cdot t_{Hetero}$ is the 2D Young's modulus (with unit [N/m]), often convenient for such atomically thin membranes. When $|V_G|$ is low, two competing effects, capacitive softening (causing a $k_{eff}$ reduction $\propto V_G^2$) and gate voltage $|V_G|$-induced tensioning (due to increased transverse deflection and thus new equilibrium position and in-plane stretching of the heterostructure membrane with increasing $|V_G|$, yielding a $k_{eff}$ addition $\propto V_G^4$), are both important but are tuning the frequency down and up, respectively. When $|V_G|$ is large enough, its tensioning of the device makes the stiffening effect dominate over the capacitive softening.

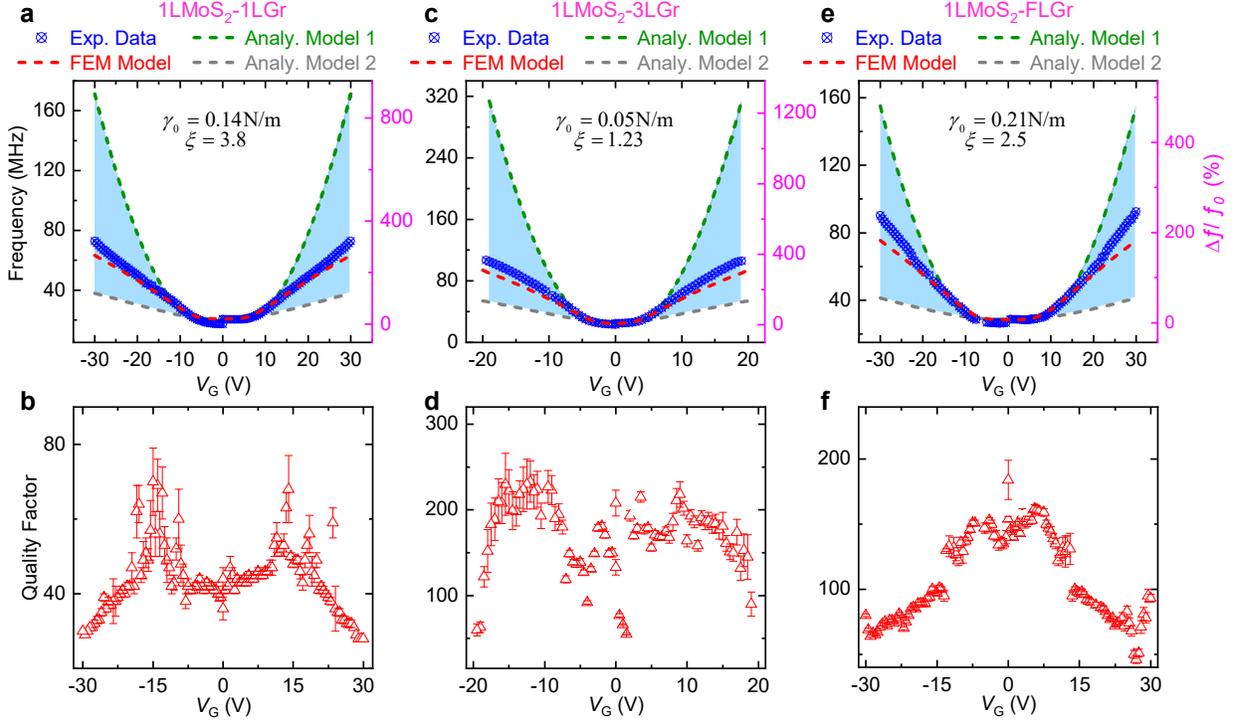

**Figure 3**: Electrostatic tuning of resonance frequencies ($f_{res}$) of atomic layer MoS$_2$-graphene heterostructure resonators – data and analysis, with $f_{res}$ versus $V_G$ in (a) 1LMoS$_2$-1LGr, (c) 1LMoS$_2$-3LGr, and (e) 1LMoS$_2$-FLGr devices, respectively. The green dashed curves and gray dashed curves are theoretical results from analytical models #1 and #2, respectively. The red dashed curves are fitting results from FEM modeling. The light blue shadow regions between model #1 and mode #2 curves are the probable tuning ranges for the van der Waals heterostructure resonators. Measured data of quality ($Q$) factors versus $V_G$: (b) 1LMoS$_2$-1LGr, (d) 1LMoS$_2$-3LGr, and (f) 1LMoS$_2$-FLGr, respectively.

As shown in Figure 3 (green dashed lines in Figure 3a,c,e), results from this analytical model #1 agree very well with the experimental data in the regime of low to moderate $|V_G|$, for all three devices, suggesting that this model nicely captures the competing softening and stiffening effects discussed above. However, it is also observed that analytical model #1 overestimates the frequency tuning in the regime of large $|V_G|$ values, which could be attributed to the deviation of mode shape and static deformation in the high $|V_G|$ regime. As $k_{eff}$ in analytical model #1 is calculated with the second derivative of all the possible elastic energy in the system, the calculated frequency tuning can therefore be regarded as an upper limit of tuning range.

To minimize the overestimation of frequency in the high $|V_G|$ regime, we explore analytical model #2 (gray dashed lines in Figure 3a,c,e). In this model, instead of considering all of the elastic energy that is stored in the system, we only count the elastic energy that corresponds to the fundamental mode resonance. Therefore, $k_{eff}$ is given as the sum of the second-order differentiation of electrostatic energy $\dfrac{\partial^2 U_{electrostatic}}{\partial z^2}$ and the second-order differentiation of elastic energy $\frac{\partial^2 \delta U_{elastic}}{\partial \delta z^2}$, where $\delta U_{elastic}$ is the elastic energy for the fundamental mode and $\delta z$ is the vibration amplitude of the device. The resonance frequency under applied voltage is given as

$$f_{res} = \frac{1}{2\pi} \sqrt{\frac{4.924 E_{Y,Hetero} t_{Hetero} \varepsilon_r - \frac{\epsilon_0 \pi a^2}{3d^3} V_G^2}{M_{eff}}}. \tag{3}$$

Here $\varepsilon_r$ is the total strain level of the vdW heterostructure membrane, which is expressed as [28]

$$\varepsilon_r = \frac{1+\varepsilon_0}{2}\sqrt{1+\frac{4z^2}{a^2}} + \frac{(1+\varepsilon_0)a}{4z^2} \ln\left(\sqrt{1+\frac{4z^2}{a^2}} + \frac{2z}{a}\right) - 1, \tag{4}$$

where $z$ is the static deflection displacement of the vdW heterostructure membrane under applied $V_G$, which is obtained by iteratively computing Eqs. (S3), (S5), and (S13) till convergence. (See the *Supporting Information* for details of analytical model #2). On the basis of the iterative calculations, an empirical relation $z \propto V_G^{4/3}$ is obtained. Considering $f_{res} \propto \varepsilon_r^{1/2}$ and $\varepsilon_r \propto z$ from Eq. (3) and Eq. (4) respectively, we expect an empirical dependency of $f_{res} \propto V_G^{2/3}$ for the stiffening contribution in Eq. (3) due to $\varepsilon_r(V_G)$. In comparison with $f_{res} \propto V_G^2$ for the stiffening term in Eq. (2) (model #1), the frequency tuning estimated using model #2 is expected to exhibit slower growing rate versus $|V_G|$.

As shown in Figure 3 (gray dashed lines in Figure 3a,c,e), model #2 also achieves very good agreement with experimental data in low $|V_G|$ regime, demonstrating model #2's effectiveness in reproducing the competing softening and stiffening characteristics in this regime. However, in contrast with model #1, model #2 exhibits an underestimation of the frequency tuning values at large $|V_G|$ because only the elastic energy induced by the oscillatory displacement ($\delta z$) is taken into account in the $k_{eff}$ and thus the $f_{res}$ calculations. Hence model #2 can be regarded as providing a lower bound of the $|V_G|$-induced frequency tuning range. Accordingly, the shadowed (light blue) region in Figure 3 between the model #1 and model #2 curves represents the probable tuning range for the vdW heterostructure resonators. Other factors that may affect the frequency tuning range include adsorbates on device surface and at the MoS$_2$-graphene interface, the vdW heterostructure's actual Young's modulus, the $|V_G|$-induced static deflection profile (at the equilibrium position) and mode shape variations, nonideal clamping conditions of the resonators, and so on, which, however, are not currently readily available to be incorporated into a single analytical model.

To better fit the experimental results, FEM modeling is applied using COMSOL Multiphysics. We utilize the *Electromechanics* module to couple the electrostatics originated from $V_G$ applied at the silicon (Si) back gate and the corresponding change in the tension in the heterostructure resonator. We use Young's modulus ($E_{Y,Gr}$=1TPa and $E_{Y,MoS2}$=0.3TPa) and thickness ($t_{1LGr}$=0.34nm and $t_{1LMoS2}$=0.7nm) of each individual constituent material. For 1LMoS$_2$-3LGr and 1LMoS$_2$-FLGr devices, clamping is considered at the edges of two materials of the heterostructure, whereas for the 1LMoS$_2$-1LGr device, the small MoS$_2$ flake is simply placed on top of the large CVD 1L graphene (MoS$_2$ not contacting the SiO$_2$). The effective initial tension of the heterostructure ($\gamma_0$ [N/m]) is considered as the summation of initial tensions in individual materials. In addition, we have introduced a mass adsorption factor ($\xi$), which takes into account the adsorbates or contaminants from the environment or fabrication process. The $\gamma_0$ and $\xi$ are treated as fitting parameters. Further, in the fitting process, we have ignored interlayer slippage between the two materials in the vdW heterostructure, as the experimental data have shown no sign of such. Compared with the two analytical models above, the COMSOL simulation results (red dashed lines along with the resultant $\gamma_0$ and $\xi$ values, in Figure 3a,c,e) show a much better agreement with the experimental results, suggesting the extracted initial tension and adsorbates factor are very close to the actual values of the devices.

In addition to ultrawide frequency tuning, another interesting observation in heterostructure resonators is the evolution of their quality ($Q$) factors versus $V_G$. Among all the heterostructure devices, the $Q$ factor first enhances and then degrades with increasing $|V_G|$ (Figure 3b,d,f). The $Q$ factors in vdW heterostructure devices during frequency tuning are influenced by several factors. First, given $Q=f/\Gamma_m$, the $Q$ factor is expected to rise with increasing frequency. Second, energy dissipation and thus damping rate $\Gamma_m$ increases with $|V_G|$ ramping, because the gap between the heterostructure membrane and Si gate becomes closer and the capacitance increases, resulting in higher oscillating current and $\Gamma_m$. Another factor is the photothermal feedback in the optical interferometry system shown in Figure 1b, where the suspended heterostructure membrane, bottom Si, and microtrench form a low-finesse optical cavity. At certain $V_G$, the condition of positive photothermal feedback is fulfilled, which compensates the dissipation of the resonator partially, leading to $Q$ enhancement [29]. The interplay among the above factors yields the measured $Q$ factor evolution. At low $|V_G|$, the increasing frequency and photothermal feedback may give rise to $Q$; at high $|V_G|$, dissipation becomes dominant and $Q$ degrades with increasing $|V_G|$.

To further explore the frequency tuning characteristics in vdW heterostructure resonators, especially to investigate its difference from single 2D material resonators, control experiments are performed. We first transfer a few-layer (4L actually) graphene sheet (**Figure 4**a) on a pre-patterned circular microcavity and measure the resonance $f_{res}$. As $V_G$ increases to 25V, $f_{res}$ of the graphene device increases from 21.6MHz to 113.6MHz, with a remarkable frequency tuning range of $\Delta f/f_0$=426%. This testifies that both single 2D material and heterostructure resonators exhibit ultrawide frequency tuning. After the frequency tuning of the graphene resonator is measured, the device bonding wire is disassembled, and another single-layer MoS$_2$ is stacked on top of graphene to form a 1LMoS$_2$-4LGr heterostructure (Figure 4b), whose gate electrode is then wire-bonded again for measurement. As $V_G$ is swept within $|V_G|$=25V, the measured $f_{res}$ of this 1LMoS$_2$-4LGr heterostructure resonator increases from 25.7MHz to 75.2MHz, showing a tuning range of $\Delta f/f_0$=193%. The ultrabroad tuning ranges and the symmetric shapes of the tuning curves in Figure 4c are consistent with those measured from other MoS$_2$-graphene heterostructure devices (Figure 3). We analyze the frequency tuning of the 4L graphene resonator and the heterostructure resonator with FEM modeling using COMSOL Multiphysics with clamping conditions that are applied to the substrate-supported regions. The FEM fitting results (red and black dashed lines in Figure 4c) agree well with the measured frequency tuning, demonstrating the fitting parameters (initial tension and adsorbates factor) are very close to the actual values in the devices. It can be also found that the frequency tuning range of the graphene device is larger than that of its corresponding vdW heterostructure device. This can be explained by the different initial tensions ($\gamma_0$=0.045N/m in graphene device and $\gamma_0$=0.16N/m in heterostructure device) and different effective masses of the graphene and vdW heterostructure devices. As shown in Figure 4d, in the low $|V_G|$ range where frequency tuning is turning from small to large, the graphene resonator's $Q$ first decreases, then increases ($Q$ vs $|V_G|$ is in "W" shape), then saturates and only fluctuates around a certain level as $|V_G|$ increases. For its 1LMoS$_2$-FLGr vdW counterpart, the $Q$ vs $|V_G|$ first takes a weak "U" shape in the low $|V_G|$ range, then slightly decreases and then slightly increases again as $|V_G|$ continues to increase. Some segments of these characteristics are similar to portions of $Q$ vs $|V_G|$ curves in Figure 3, but specific $|V_G|$ ranges vary, and the variations of $Q$ vs $|V_G|$ are smaller than in Figure 3. These suggest detailed differences in the relative importance of the competing factors that affect $Q$ and their $|V_G|$ dependency.

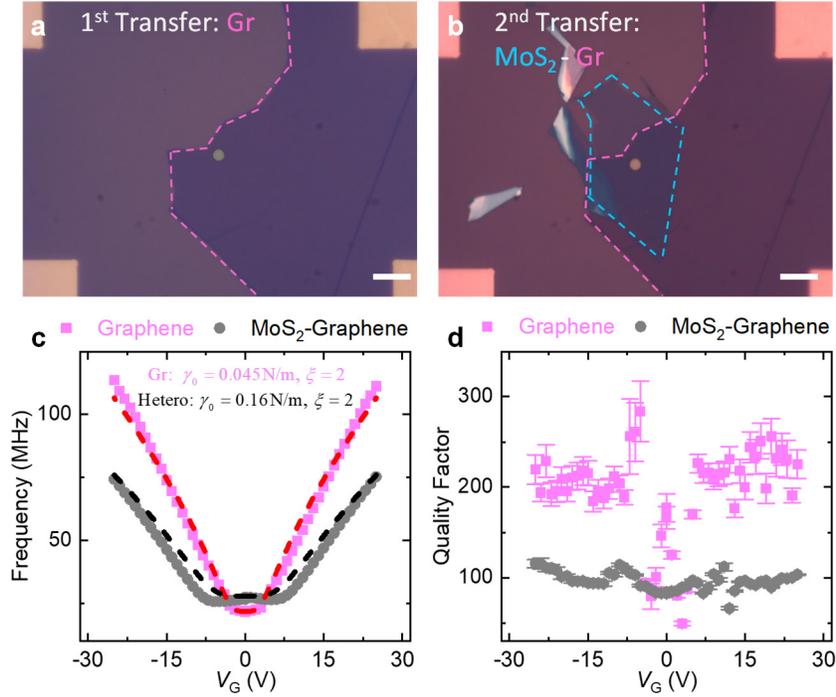

**Figure 4**: Control experiment for frequency tuning in graphene and MoS$_2$-graphene heterostructure resonators. Optical images of (a) a few-layer graphene (FLGr) resonator and (b) a 1LMoS$_2$-FLGr heterostructure resonator with single-layer (1L) MoS$_2$ on top of the FL graphene in (a). (c) Frequency tuning and (d) quality ($Q$) factor evolution of the device with increasing $|V_G|$ before and after the 1LMoS$_2$ is stacked atop graphene. The dashed lines in panel (c) show FEM modeling results, with $\gamma_0$=0.045N/m and $\xi$=2 for the FLGr device and $\gamma_0$=0.16N/m and $\xi$=2 for the 1LMoS$_2$-FLGr heterostructure device. *Scale bars*: 10μm.

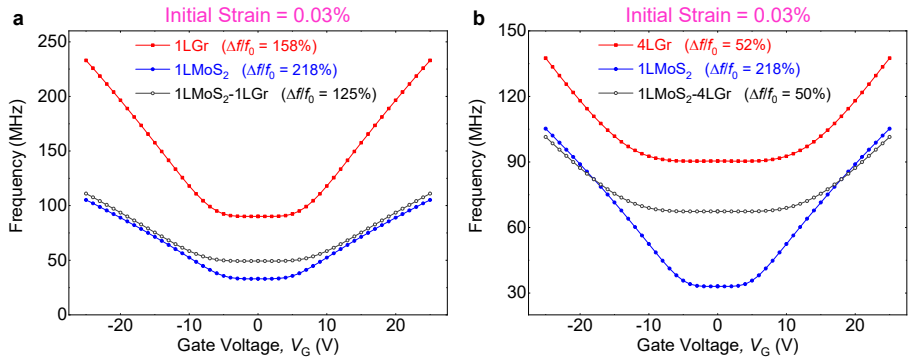

**Figure 5**: Computed frequency tuning ranges of single-material and bi-material heterostructure resonators. (a) Comparison of $f_{res}$ tuning among 1LMoS$_2$, 1LGr, and 1LMoS$_2$-1LGr resonators with a fixed initial strain of $\varepsilon_0$ = 0.03%. (b) Comparison of $f_{res}$ tuning among 1LMoS$_2$, 4LGr, and 1LMoS$_2$-4LGr resonators with a fixed initial strain of $\varepsilon_0$ = 0.03%.

It should be pointed out that in the control experiment, because the initial strains of graphene and heterostructure resonators are not identical, the comparison does not represent the ideal scenario. In fact, after additive stacking of the 1LMoS$_2$, the heterostructure resonator has a much higher initial strain, which contributes to a comparatively smaller tuning range. To gain further insights into the difference between single-material and vdW heterostructure resonators, we simulate the $f_{res}$ tuning of 1LGr, 1LMoS$_2$ and 1LMoS$_2$-1LGr devices (of the same size) with the same initial strain of $\varepsilon_0 = 0.03\%$ via FEM modeling. We find that the 1LMoS$_2$ resonator offers the largest tuning range ($\Delta f/f_0 \approx 218\%$) among these three resonators, and the heterostructure resonator has the lowest tuning range ($\Delta f/f_0 \approx 125\%$) (**Figure 5**a). In addition, we also simulate the frequency tuning of 4LGr and 1LMoS$_2$-4LGr (the same as in the control experiment) with the same initial strain of $\varepsilon_0 = 0.03\%$, which also demonstrates that the 1LMoS$_2$-4LGr device shows the lowest tuning range ($\Delta f/f_0 \approx 50\%$) (**Figure 5**b). These simulations demonstrate that given the same initial strain, the material with the lowest $E_{Y,2D}$ (= $E_Y \cdot t$, with unit [N/m]) will provide the highest $f_{res}$ tuning range.

## Conclusions

In summary, we have demonstrated for the first time the ultrawide electrostatic frequency tuning of MoS$_2$-graphene vdW heterostructure nanoelectromechanical resonators. All the tested devices show frequency tuning range $\Delta f/f_0 > 200\%$, up to 370%, which is much wider than attained by traditional bimorph N/MEMS resonators and even some resonators made of single 2D materials. Two analytical models and FEM modeling have been used to simulate the frequency tuning responses of the heterostructure resonators. We have shown that lower 2D Young's modulus ($E_{Y,2D} = E_Y \cdot t$) and initial strain lead to higher frequency tuning range. This study not only demonstrates 2D vdW heterostructure resonators with ultrawide frequency tuning, but also paves the way for building highly tunable 2D NEMS oscillators and ultrasensitive atomically thin heterostructure transducers and signal processing elements.

**Supporting Information**

The Supporting Information is available at the end of this document.

    Experimental methods; Analytical model #1 and analytical model #2 of heterostructure resonator frequency tuning; Benchmarking of frequency tuning (PDF)

**Notes**

The authors declare no competing financial interest.

### ACKNOWLEDGEMENTS

We thank the support from National Science Foundation CAREER Award (Grant ECCS-1454570, 2015708) and CCSS Award (Grant ECCS-1509721). Part of the device fabrication was performed at the Cornell Nanoscale Science and Technology Facility (CNF), a member of the National Nanotechnology Coordinated Infrastructure (NNCI), supported by the National Science Foundation (Grant ECCS-1542081). Dr. Teng Zhang acknowledges the start-up funds from Syracuse University.

# References

1 Feng, X. L.; White, C. J.; Hajimiri, A.; Roukes, M. L. A Self-Sustaining Ultrahigh-Frequency Nanoelectromechanical Oscillator. *Nat. Nanotechnol.* **2008**, *3* (6), 342–346.

2 Chen, C.; Lee, S.; Deshpande, V. V; Lee, G.-H.; Lekas, M.; Shepard, K.; Hone, J. Graphene Mechanical Oscillators with Tunable Frequency. *Nat. Nanotechnol.* **2013**, *8* (12), 923–927.

3 Nguyen, C. T.-C. MEMS Technology for Timing and Frequency Control. *IEEE Trans. Ultrason. Ferroelect. Freq. Contr*. **2007**, *54* (2), 251-270.

4 Kozinsky, I.; Postma, H. W. C.; Bargatin, I.; Roukes, M. L. Tuning Nonlinearity, Dynamic Range, and Frequency of Nanomechanical Resonators. *Appl. Phys. Lett.* **2006**, *88* (25), 253101.

5 Ye, F.; Lee, J.; Feng, P. X.-L. Electrothermally Tunable Graphene Resonators Operating at Very High Temperature up to 1200 K. *Nano Lett.* **2018**, *18* (3), 1678–1685.

6 Jun, S. C.; Huang, X. M. H.; Manolidis, M.; Zorman, C. A.; Mehregany, M.; Hone, J. Electrothermal Tuning of Al–SiC Nanomechanical Resonators. *Nanotechnology* **2006**, *17* (5), 1506–1511.

7 Karabalin, R. B.; Feng, X. L.; Roukes, M. L. Parametric Nanomechanical Amplification at Very High Frequency. *Nano Lett.* **2009**, *9*, 3116–3123.

8 Piazza, G.; Abdolvand, R.; Ayazi, F. Voltage-Tunable Piezoelectrically-Transduced Single-Crystal Silicon Resonators on SOI Substrate. In *The 16th International Conference on Micro Electro Mechanical Systems, 2003. MEMS-03 Kyoto. IEEE*; **2003**; pp 149–152.

9 Svilicic, B.; Mastropaolo, E.; Flynn, B.; Cheung, R. Electrothermally Actuated and Piezoelectrically Sensed Silicon Carbide Tunable MEMS Resonator. *IEEE Electron Device Lett.* **2012**, *33* (2), 278–280.

10 Novoselov, K. S.; Geim, A. K.; Morozov, S. V; Jiang, D.; Zhang, Y.; Dubonos, S. V.; Grigorieva, I. V; Firsov, A. A. Electric Field Effect in Atomically Thin Carbon Films *Science* **2004**, *306* (5696), 666–669.

11 Mak, K. F.; Lee, C.; Hone, J.; Shan, J.; Heinz, T. F. Atomically Thin $MoS_2$: A New Direct-Gap Semiconductor. *Phys. Rev. Lett.* **2010**, *105* (13), 136805.

12 Geim, A. K.; Grigorieva, I. V. van der Waals Heterostructures. *Nature* **2013**, *499*, 419.

13 Ye, F.; Lee, J.; Feng, P. X.-L. Atomic Layer $MoS_2$-Graphene van der Waals Heterostructure Nanomechanical Resonators. *Nanoscale* **2017**, *9* (46), 18208–18215.

14 Kim, S.; Yu, J.; van der Zande, A. M. Nano-electromechanical Drumhead Resonators from Two-Dimensional Material Bimorphs. *Nano Lett.* **2018**, *18* (11), 6686–6695.

15 Kumar, R.; Session, D. W.; Tsuchikawa, R.; Homer, M.; Paas, H.; Watanabe, K.; Taniguchi, T.; Deshpande, V. V. Circular Electromechanical Resonators Based on Hexagonal-Boron Nitride-Graphene Heterostructures. *Appl. Phys. Lett.* **2020**, *117* (18), 183103.

16 Yang, R., Zheng, X., Wang Z., Miller, C. J. and Feng, P. X.-L. Multilayer $MoS_2$ Transistors Enabled by a Facile Dry-Transfer Technique and Thermal Annealing. *J. Vac. Sci. Technol. B, Nanotechnol. Microelectron. Mater. Process. Meas. Phenom.* **2014**, *32* (6), 061203

17 Chen, C.; Rosenblatt, S.; Bolotin, K. I.; Klab, W.; Kim, P.; Kymissis, I.; Stormer, H. L.; Heinz, T. F.; Hone, J. Performance of Monolayer Graphene Nanomechanical Resonators with Electrical Readout. *Nat. Nanotechnol.* **2009**, *4* (12), 861–867.

18 Reserbat-Plantey, A.; Marty, L.; Arcizet, O.; Bendiab, N.; Bouchiat, V. A Local Optical Probe for Measuring Motion and Stress in a Nanoelectromechanical System. *Nat. Nanotechnol.* **2012**, *7* (3), 151–155.

19  Miao, T.; Yeom, S.; Wang, P.; Standley, B.; Bockrath, M. Graphene Nanoelectromechanical Systems as Stochastic-Frequency Oscillators. *Nano Lett.* **2014**, *14* (6), 2982–2987.

20  Weber, P.; Güttinger, J.; Tsioutsios, I.; Chang, D. E.; Bachtold, A. Coupling Graphene Mechanical Resonators to Superconducting Microwave Cavities. *Nano Lett.* **2014**, *14* (5), 2854–2860.

21  Song, X.; Oksanen, M.; Sillanpää, M. A.; Craighead, H. G.; Parpia, J. M.; Hakonen, P. J. Stamp Transferred Suspended Graphene Mechanical Resonators for Radio Frequency Electrical Readout. *Nano Lett.* **2011**, *12* (1), 198–202

22  Li, H.; Zhang, Q.; Yap, C. C. R.; Tay, B. K.; Edwin, T. H. T.; Olivier, A.; Baillargeat, D. From Bulk to Monolayer $MoS_2$: Evolution of Raman Scattering. *Adv. Funct. Mater.* **2012**, *22* (7), 1385–1390.

23  Ferrari, A. C.; Meyer, J. C.; Scardaci, V.; Casiraghi, C.; Lazzeri, M.; Mauri, F.; Piscanec, S.; Jiang, D.; Novoselov, K. S.; Roth, S.; Geim, A. K. Raman Spectrum of Graphene and Graphene Layers. *Phys. Rev. Lett.* **2006**, *97* (18), 187401.

24  Witkamp, B.; Poot, M.; van der Zant, H. S. J. Bending-Mode Vibration of a Suspended Nanotube Resonator. *Nano Lett.* **2006**, *6* (12), 2904–2908.

25  Elsayed, M. Y.; Cicek, P.; Nabki, F.; El-Gamal, M. N. Bulk Mode Disk Resonator with Transverse Piezoelectric Actuation and Electrostatic Tuning. *J. Microelectromechanical Syst.* **2016**, *25* (2), 252–261.

26  Sazonova, V.; Yaish, Y.; Üstünel, H.; Roundy, D.; Arias, T. A.; McEuen, P. L. A Tunable Carbon Nanotube Electromechanical Oscillator. *Nature* **2004**, *431* (7006), 284–287.

27  Lee, J.; Wang, Z.; He, K.; Yang, R.; Shan, J.; Feng, P. X.-L. Electrically Tunable Single- and Few-Layer $MoS_2$ Nanoelectromechanical Systems with Broad Dynamic Range. *Sci. Adv.* **2018**, *4* (3), eaao6653.

28  Mei, T.; Lee, J.; Xu, Y.; Feng, P. X.-L. Frequency Tuning of Graphene Nanoelectromechanical Resonators via Electrostatic Gating. *Micromachines*, **2018**, *9* (6), 312.

29  Barton, R. A; Storch, I. R.; Adiga, V. P.; Sakakibara, R.; Cipriany, B. R.; Ilic, B.; Wang, S. P.; Ong, P.; McEuen, P. L.; Parpia, J. M.; Craighead, H. G. Photothermal Self-Oscillation and Laser Cooling of Graphene Optomechanical Systems. *Nano Lett.* **2012**, *12* (9), 4681–4686.

# Ultrawide Frequency Tuning of Atomic Layer van der Waals Heterostructure Electromechanical Resonators

Fan Ye[1†], Arnob Islam[1†], Teng Zhang[3,4], Philip X.-L. Feng[1,2*]

[1]*Department of Electrical, Computer, & Systems Engineering, Case School of Engineering, Case Western Reserve University, Cleveland, OH 44106, USA*

[2]*Department of Electrical & Computer Engineering, Herbert Wertheim College of Engineering, University of Florida, Gainesville, FL 32611, USA*

[3]*Department of Mechanical & Aerospace Engineering, and* [4]*BioInspired Syracuse, Syracuse University, Syracuse, NY 13244, USA*

**Table of Contents**

[†]Equally Contributed Authors.　[*]Corresponding Author.　Email: philip.feng@ufl.edu.

## S1. Experimental Methods

### Fabrication of MoS₂-Graphene Heterostructure Resonators

The graphene flakes are exfoliated from graphite crystal onto PDMS stamps and then transferred onto pre-patterned microcavities on 290nm-thick SiO$_2$ on Si substrate using the all-dry transfer techniques [S1]. The diameter of the pre-patterned circular microcavity is ~3.2μm and depth is 290nm. After graphene transfer, a MoS$_2$ flake is exfoliated, carefully selected and then stacked on top of graphene flake using the same all-dry transfer techniques with alignment [S1]. CVD graphene layers are employed in 1LMoS$_2$-1LGraphene (Gr) devices, while exfoliated graphene flakes are used in making 1LMoS$_2$-3LGr and 1L MoS$_2$-FLGr devices.

### Electrostatic Excitation and Interferometric Measurement of Heterostructure Resonators

The resonance characteristics of the MoS$_2$-graphene heterostructure devices are measured using an ultrasensitive optical interferometry system (Figure 1b). The resonance motion is electrostatically excited by applying a radio frequency (RF) signal $v_{RF}$ to the gate. In addition to $v_{RF}$, a DC polarization voltage $V_G$ is applied to provide electrostatic gate tuning of the heterostructure membrane. We employ a network analyzer to provide the $v_{RF}$ drive signal, sweeping from 1MHz to 150MHz. The resonance motion is detected by a 633nm red laser (spot size ~1μm) with power below 300μW to avoid laser heating. The measured resonance response signal from the photodetector in the frequency domain is recorded by the same network analyzer.

### Raman Scattering and Photoluminescence (PL) Measurement

MoS$_2$-graphene heterostructure devices are preserved in a vacuum chamber and measured using a customized micro-Raman system that is integrated into the optical interferometric resonance measurement system (see the section above). During the measurement, we focus a 532nm green laser at the center of the heterostructure device. The spot size of laser around ~1μm and laser power is below 200μW to avoid excessive laser heating. Raman scattered light from the sample is collected in backscattering geometry and then guided to a spectrometer (Horiba iHR550) with a 2400g/mm grating. Raman signal is recorded using a liquid-nitrogen-cooled CCD. PL spectra are recorded using the same spectrometer and CCD with a 1200g/mm grating. The typical integration time is 10s for PL and 60s for Raman measurement.

## S2. Analytical Model #1 of Heterostructure Resonator Frequency Tuning

The heterostructure circular drumhead resonator's mechanical response under capacitive coupling can be analyzed by considering both elastic energy in the suspended membrane and electrostatic energy stored in the capacitor formed by the 2D heterostructure and its gate.

The effective Young modulus of heterostructure is

$$E_{Y,Hetero} = \frac{E_{Y,MoS_2} t_{MoS_2} + E_{Y,Gr} t_{Gr}}{t_{Hetero}}, \tag{S1}$$

where $E_{Y,MoS_2}$ and $E_{Y,Gr}$ are the Young's moduli [N/m$^2$] of MoS$_2$ and graphene, respectively; and $t_{MoS_2}$ and $t_{Gr}$ refer to the thicknesses of MoS$_2$ and graphene, respectively; and $t_{Hetero} = t_{MoS_2} + t_{Gr}$. The elastic energy of the heterostructure resonator is a function of its displacement $W(r,\theta)$:

$$U_{elastic} = \int_0^{2\pi} d\theta \int_0^a rdr \left\{ \frac{E_{Y,Hetero} t_{Hetero}}{2} \left[ \varepsilon_{0,r} \left( \frac{\partial W}{\partial r} \right)^2 + \frac{\varepsilon_{0,\theta}}{r^2} \left( \frac{\partial W}{\partial \theta} \right)^2 \right] + \frac{t_{Hetero}}{8} \frac{E_{Y,Hetero} t_{Hetero}}{1-v^2} \left( \frac{\partial W}{\partial r} \right)^2 \right\}, \quad (S2)$$

where $E_Y$, $v$, $\varepsilon_{0,r}$, $\varepsilon_{0,\theta}$, $t$, $a$ are Young's modulus (with unit [N/m$^2$] or [Pa]), Poisson's ratio, initial radial strain, initial tangential strain, thickness, and device radius, respectively. By assuming the radial profile of the heterostructure membrane stretched by the electrostatic force is parabolic, $W(r,\theta) = z(1-r^2/a^2)$, where $z$ is the deflection at its center, Eq. (S2) can be rewritten as

$$U_{elastic} = \frac{\pi E_{Y,Hetero} t}{1-v^2} \left[ \frac{2z^4}{3a^2} + (1-v^2) \varepsilon_{0,r} z^2 + \frac{1}{2} \varepsilon_{0,r} a^2 \right]. \quad (S3)$$

The electrostatic potential energy induced by the applied gate voltage $V_G$ is:

$$U_{electrostatic} = -\frac{1}{2} C_g V_G^2, \quad (S4)$$

where $C_g$ is capacitance of the capacitor formed by bottom Si gate and heterostructure membrane. Under any given gate polarization voltage $V_G$, the electrostatic energy is changed when the heterostructure membrane moves toward the gate and $C_g$ can be approximated as

$$C_g = \int \frac{\epsilon_0}{d - z(1-r^2/a^2)} rdrd\theta, \quad (S5)$$

which can be further simplified in terms of the static deflection $z$,

$$C_g = \frac{\epsilon_0 A}{d} + \frac{\epsilon_0 A}{2d^2} z + \frac{\epsilon_0 A}{3d^3} z^2, \quad (S6)$$

where $\epsilon_0$ is the vacuum permittivity, $d$ is the vacuum cavity depth (or gap), and $A$ is the area of the device. The equilibrium position of static displacement $z$ can be calculated by minimizing the total energy (sum of elastic energy and electrostatic energy):

$$F_{tot} = -\frac{\partial (U_{elastic} + U_{electrostatic})}{\partial z}$$
$$= \frac{8\pi E_{Y,Hetero} t_{Hetero}}{3(1-v^2)a^2} z^3 + \left( 2\pi E_{Y,Hetero} t_{Hetero} \varepsilon_0 - \frac{1}{2} \frac{\partial^2 C_g}{\partial z^2} V_G^2 \right) z - \frac{1}{2} \frac{\partial C_g}{\partial z} V_G^2 = 0 \quad (S7)$$

Neglecting the first term in Eq. (S7) as $z^3 \to 0$ and assuming $\frac{\partial^2 C_g}{\partial z^2} \approx 0$, we obtain

$$z \approx \frac{\epsilon_0}{8 E_{Y,Hetero} t_{Hetero} \varepsilon_0} \frac{a^2}{d^2} V_G^2.$$

The resonance frequency of the heterostructure resonator is given by $f_{res} = \sqrt{k_{eff}/M_{eff}}/2\pi$. Here, the effective spring constant of the device $k_{eff}$ can be obtained by second order differentiation of the total potential energy. The $k_{eff}$ is given by

$$k_{eff} = \frac{\partial^2 (U_{elastic} + U_{electrostatic})}{\partial z^2} = \frac{8\pi E_{Y,Hetero} t_{Hetero}}{(1-v^2)a^2} z^2 + 4.9 E_{Y,Hetero} t_{Hetero} \varepsilon_0 - \frac{1}{2} \frac{\partial^2 C_g}{\partial z^2} V_G^2. \quad (S8)$$

By substituting $z$ and $\frac{\partial^2 C_g}{\partial z^2}$ into Eq. (S8), we have

$$k_{\text{eff}} = 4.9 E_{Y,\text{Hetero}} t_{\text{Hetero}} \varepsilon_0 - \frac{\epsilon_0 \pi a^2}{3d^3} V_G^2 + \frac{\pi \epsilon_0^2}{8(1-v^2) E_{Y,\text{Hetero}} t_{\text{Hetero}} \varepsilon_0^2} \frac{a^2}{d^4} V_G^4. \quad (S9)$$

Therefore the resonance frequency tuning by gate voltage $V_G$ can then be expressed as

$$f_{\text{res}} = \frac{1}{2\pi} \sqrt{\frac{4.9 E_{Y,\text{Hetero}} t_{\text{Hetero}} \varepsilon_0 - \frac{\epsilon_0 \pi a^2}{3d^3} V_G^2 + \frac{\pi \epsilon_0^2}{8(1-v^2) E_{Y,\text{Hetero}} t_{\text{Hetero}} \varepsilon_0^2} \frac{a^2}{d^4} V_G^4}{M_{\text{eff}}}}. \quad (S10)$$

## S3. Analytical Model #2 of Heterostructure Resonator Frequency Tuning

In this model, instead of considering all the elastic energy that is stored in the device, we only count the part that is for the fundamental mode resonance. Thus the effective spring constant is given as

$$k_{\text{eff}} = \frac{\partial^2 \delta U_{\text{elastic}}}{\partial \delta z^2} + \frac{\partial^2 U_{\text{electrostatic}}}{\partial z^2}, \quad (S11)$$

where $\delta U_{\text{elastic}}$ is the elastic energy for the fundamental mode and $\delta z$ is the vibration amplitude of the device. By inserting the mode shape of the circular membrane in its fundamental resonance mode, $W_{\text{mode}} = \delta z J_0 \left( \beta_{01} \cdot \frac{r}{a} \right)$ ($J_0$: the 0$^{\text{th}}$-order Bessel function $J$; $\beta_{01} = 2.405$), into Eq. (S7), the elastic energy at resonance is given as

$$\delta U_{\text{elastic}} = 0.271 \frac{2.405^2 \pi E_{Y,\text{Hetero}} t_{\text{Hetero}} \varepsilon_r}{2} (\delta z)^2 + \frac{\pi E_{Y,\text{Hetero}} t_{\text{Hetero}} \varepsilon_r a^2}{2(1-v^2)}, \quad (S12)$$

where $\varepsilon_r$ is the total strain in the 2D drumhead due to radial elongation of the membrane and is estimated as follows

$$\varepsilon_r = \frac{1+\varepsilon_0}{2a} \int_{-a}^{a} \sqrt{1 + \left(\frac{\partial u}{\partial r}\right)^2} dr - 1 = \frac{1+\varepsilon_0}{2} \sqrt{1 + \frac{4z^2}{a^2}} + \frac{(1+\varepsilon_0)a}{4z^2} \ln\left(\sqrt{1 + \frac{4z^2}{a^2}} + \frac{2z}{a}\right) - 1, \quad (S13)$$

where $z$ stands for the equilibrium displacement of the van der Waals heterostructure membrane under applied gate voltage $V_G$, which is obtained by iteratively calculating Eq. (S3), Eq. (S5) and Eq. (S13) until reaching convergence. By inserting Eq. (S11) and Eq. (S12) into $f_{\text{res}} = \sqrt{k_{\text{eff}}/M_{\text{eff}}}/2\pi$, we can obtain

$$f_{\text{res}} = \frac{1}{2\pi} \sqrt{\frac{4.924 E_{Y,\text{Hetero}} t_{\text{Hetero}} \varepsilon_r - \frac{\varepsilon_0 \pi a^2}{3d^3} V_G^2}{M_{\text{eff}}}}. \quad (S14)$$

## S4. Benchmarking of Frequency Tuning

Figure S1 compares the figure of merit (FoM) of normalized frequency tuning – frequency tuning per voltage applied ($\frac{\Delta f / f_0}{V}$) in this work and other 2D resonators reported in literature. It can be clearly seen that the frequency tuning FoM in this work is much larger than those reported in other 2D resonators.

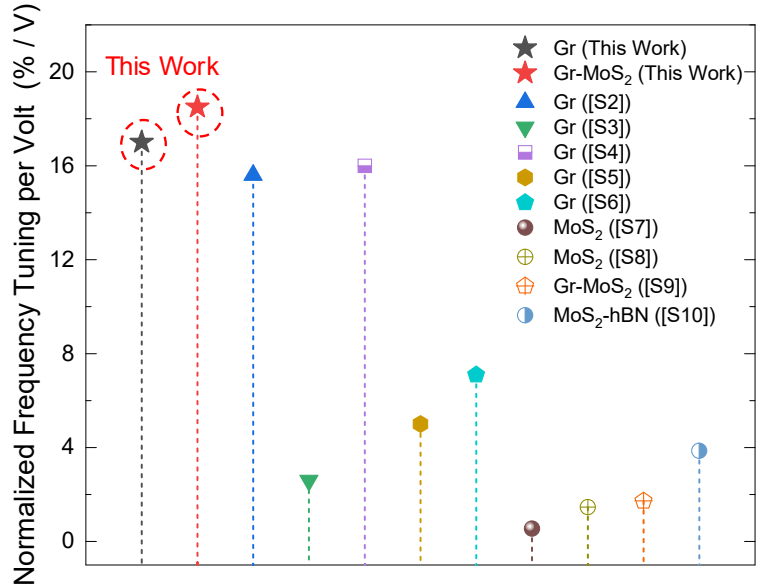

**Figure S1**: Benchmarking of frequency tuning capability in graphene and $MoS_2$-graphene van der Waals heterostructure resonators: Comparison of voltage-normalized frequency tuning between devices in this work and other 2D NEMS resonators reported in literature [S2,S3,S4,S5,S6,S7,S8,S9,S10].

Figure S2 compares the FoM, frequency tuning per voltage applied ($\frac{\Delta f / f_0}{V}$), measured in this work and data from other representative tunable N/MEMS resonators in literature. It can be clearly seen that the frequency tuning FoM in this work is much greater than in other N/MEMS devices.

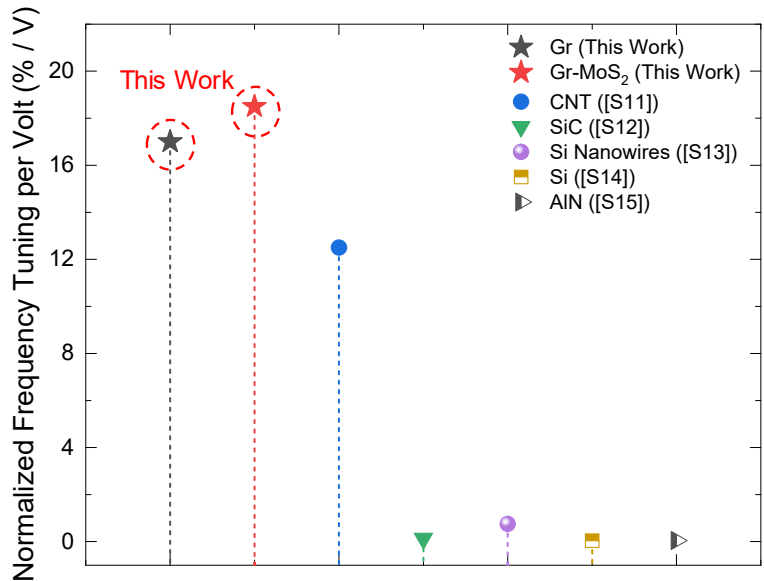

**Figure S2**: Benchmark of frequency tuning capability in graphene and $MoS_2$-graphene van der Waals heterostructure resonators: Comparison of the voltage-normalized frequency tuning between devices in this work and other N/MEMS resonators in literature [S11,S12,S13,S14,S15].

# References

S1  Yang, R.; Zheng, X.; Wang Z.; Miller, C. J.; and Feng, P. X.-L. Multilayer MoS$_2$ Transistors Enabled by a Facile Dry-Transfer Technique and Thermal Annealing. *J. Vac. Sci. Technol. B, Nanotechnol. Microelectron. Mater. Process. Meas. Phenom.* **2014**, *32* (6), 061203.

S2  Chen, C.; Rosenblatt, S.; Bolotin, K. I.; Kalb, W.; Kim, P.; Kymissis, I.; Stormer, H. L.; Heinz, T. F.; Hone, J. Performance of Monolayer Graphene Nanomechanical Resonators with Electrical Readout. *Nat. Nanotechnol.* **2009**, *4* (12), 861–867.

S3  Singh, V.; Sengupta, S.; Solanki, H. S.; Dhall, R.; Allain, A.; Dhara, S.; Pant, P.; & Deshmukh, M. M. Probing Thermal Expansion of Graphene and Modal Dispersion at Low-Temperature Using Graphene Nanoelectromechanical Systems Resonators. *Nanotechnology* **2010**, *21* (16), 165204.

S4  van der Zande, A. M.; Barton, R. A.; Alden, J. S.; Ruiz-Vargas, C. S.; Whitney, W. S.; Pham, P. H. Q.; Park, J.; Parpia, J. M.; Craighead, H. G.; McEuen, P. L. Large-Scale Arrays of Single-Layer Graphene Resonators. *Nano Lett.* **2010**, *10*, 4869–4873.

S5  Barton, R. A.; Storch, I. R.; Adiga, V. P.; Sakakibara, R.; Cipriany, B. R.; Ilic, B.; Wang, S. P.; Ong, P.; McEuen, P. L.; Parpia, J. M.; Craighead, H. G. Photothermal Self-Oscillation and Laser Cooling of Graphene Optomechanical Systems. *Nano Lett.* **2012**, *12* (9), 4681–4686.

S6  Song, X.; Oksanen, M.; Sillanpää, M. A.; Craighead, H. G.; Parpia, J. M.; Hakonen, P. J. Stamp Transferred Suspended Graphene Mechanical Resonators for Radio Frequency Electrical Readout. *Nano Lett.* **2012**, *12* (1), 198–202.

S7  Guan, F.; Kumaravadivel, P.; Averin, D. V; Du, X. Tuning Strain in Flexible Graphene Nanoelectromechanical Resonators. *Appl. Phys. Lett.* **2015**, *107* (19), 193102.

S8  Lee, J.; Wang, Z.; He, K.; Yang, R.; Shan, J.; Feng, P. X.-L. Electrically Tunable Single- and Few-Layer MoS$_2$ Nanoelectromechanical Systems with Broad Dynamic Range. *Sci. Adv.* **2018**, *4* (3), eaao6653.

S9  Kim, S.; Yu, J.; van der Zande, A. M. Nano-Electromechanical Drumhead Resonators from Two-Dimensional Material Bimorphs. *Nano Lett.* **2018**, *18* (11), 6686–6695.

S10  Kumar, R.; Session, D. W.; Tsuchikawa, R.; Homer, M.; Paas, H.; Watanabe, K.; Taniguchi, T.; Deshpande, V. V. Circular Electromechanical Resonators Based on Hexagonal-Boron Nitride-Graphene Heterostructures. *Appl. Phys. Lett.* **2020**, *117* (18), 183103.

S11  Witkamp, B.; Poot, M.; van der Zant, H. S. J. Bending-Mode Vibration of a Suspended Nanotube Resonator. *Nano Lett.* **2006**, *6* (12), 2904–2908.

S12  Kozinsky, I.; Postma, H. W. C.; Bargatin, I.; Roukes, M. L. Tuning Nonlinearity, Dynamic Range, and Frequency of Nanomechanical Resonators. *Appl. Phys. Lett.* **2006**, *88* (25), 253101.

S13  He, R.; Feng, X. L.; Roukes, M. L.; Yang, P. Self-Transducing Silicon Nanowire Electromechanical Systems at Room Temperature. *Nano Lett.* **2008**, *8* (6), 1756–1761.

S14  Piazza, G.; Abdolvand, R.; Ayazi, F. Voltage-Tunable Piezoelectrically-Transduced Single-Crystal Silicon Resonators on SOI Substrate. In *The 16$^{th}$ Annual International Conference on Micro Electro Mechanical Systems (MEMS'03)*, Kyoto, Japan, January 19-23, **2003**, pp 149–152.

S15  Karabalin, R. B.; Feng, X. L.; Roukes, M. L. Parametric Nanomechanical Amplification at Very High Frequency. *Nano Lett.* **2009**, *9* (9), 3116–3123.